\PassOptionsToPackage{table}{xcolor}

\documentclass[sigconf]{acmart}
\usepackage{conf26-gen-topic}

\begin{document}
\begin{acronym}

	\acro{DL20}[DL20]{TREC Deep Learning Track 2020}
	\acro{DL19}[DL19]{TREC Deep Learning Track 2019}
	\acro{R04}[R04]{TREC Robust Track 2004}

\end{acronym}
\title[Formalized Information Needs Improve Large-Language-Model Relevance Judgments]{Formalized Information Needs Improve Large-Language-Model~Relevance~Judgments}

\settopmatter{authorsperrow=4,printacmref=true}

\newcommand{\colognInst}{TH Köln}

\author{Jüri Keller}
\orcid{0000-0002-9392-8646}
\affiliation{%
	\institution{\colognInst}
	\city{Cologne}
	\country{Germany}
}
\author{Maik Fröbe}
\orcid{0000-0002-1003-981X}
\affiliation{%
	\institution{Friedrich-Schiller-\\Universität Jena}
	\city{Jena}
	\country{Germany}
}
\author{Björn Engelmann}
\orcid{0009-0000-7074-9066}
\affiliation{%
	\institution{\colognInst}
	\city{Cologne}
	\country{Germany}
}
\author{Fabian Haak}
\orcid{0000-0002-3392-7860}
\affiliation{%
	\institution{\colognInst}
	\city{Cologne}
	\country{Germany}
}
\author{Timo Breuer}
\orcid{0000-0002-1765-2449}
\affiliation{%
	\institution{\colognInst}
	\city{Cologne}
	\country{Germany}
}
\author{Birger Larsen}
\orcid{0000-0002-3622-2698}
\affiliation{%
	\institution{Aalborg University}
	\city{Copenhagen}
	\country{Denmark}
}
\author{Philipp Schaer}
\orcid{0000-0002-8817-4632}
\affiliation{%
	\institution{\colognInst}
	\city{Cologne}
	\country{Germany}
}

\renewcommand{\shortauthors}{Jüri Keller et al.}

\copyrightyear{2026}
\acmYear{2026}
\setcopyright{cc}
\setcctype{by}
\acmConference[SIGIR '26]{Proceedings of the 49th International ACM SIGIR Conference on Research and Development in Information Retrieval}{July 20--24, 2026}{Melbourne, VIC, Australia}
\acmBooktitle{Proceedings of the 49th International ACM SIGIR Conference on Research and Development in Information Retrieval (SIGIR '26), July 20--24, 2026, Melbourne, VIC, Australia}
\acmDOI{10.1145/3805712.3809561}
\acmISBN{979-8-4007-2599-9/2026/07}

\begin{abstract}
Cranfield-style retrieval evaluations with too few or too many relevant documents or with low inter-assessor agreement on relevance can reduce the reliability of observations.
In evaluations with human assessors, information needs are often formalized as retrieval topics to avoid an excessive number of relevant documents while maintaining good agreement.
However, emerging evaluation setups that use Large Language Models (LLMs) as relevance assessors often use only queries, potentially decreasing the reliability.
To study whether LLM relevance assessors benefit from formalized information needs, we synthetically formalize information needs with LLMs into topics that follow the established structure from previous human relevance assessments (i.e., descriptions and narratives).
We compare assessors using synthetically formalized topics against the LLM-default query-only assessor on the~2019/2020~editions of TREC Deep Learning and Robust04.
We find that assessors without formalization judge many more documents relevant and have a lower agreement, leading to reduced reliability in retrieval evaluations.
Furthermore, we show that the formalized topics improve agreement between human and LLM relevance judgments, even when the topics are not highly similar to their human counterparts.
Our findings indicate that LLM relevance assessors should use formalized information needs, as is standard for human assessment, and synthetically formalize topics when no human formalization exists to improve evaluation reliability.

\end{abstract}



\maketitle

\begin{table}[h]
    \centering
    \small 
    \caption{Comparison of the traditional TREC-style relevance assessment (where topics are formalized) with LLM relevance assessment when no formalization exists and judges then by default only use the query without formalization. We propose to formalize the description and narratives with LLMs when no human formalization exists and show advantages~($+$) and disadvantages~($-$) of the different methodologies on the costs, the reliability (Rel.), and bias towards positivity (Pos.).}
    \label{tab:comparison}
    
    \setlength{\tabcolsep}{4.4pt} 
    
    \begin{tabular}{@{}l ccc c c c w{c}{0.9cm}@{}}
        \toprule
        & \multicolumn{3}{c}{\textbf{Topic Formalization}} & \textbf{Judge} & \multicolumn{3}{c}{\textbf{Resulting Judgments}} \\
        \cmidrule(lr){2-4} \cmidrule(l){6-8}
        \textbf{Method} & \textbf{Qry.} & \textbf{Desc.} & \textbf{Narr.} &  & \textbf{Cost} & \textbf{Rel.} & \textbf{Pos.} \\
        \midrule
        TREC    & \small\faUser\normalsize & \small\faUser\normalsize & \small\faUser\normalsize & \small\faUser\normalsize 
                & \footnotesize $---$ & \footnotesize $+++$ & \footnotesize Low \\
                \midrule
        
        LLM-Default & \small\faUser\normalsize & --- & --- & \small\faRobot\normalsize 
                & \footnotesize $++$ & \footnotesize $+$ & \footnotesize High \\
        
        \textbf{LLM-Ours} & \small\faUser\normalsize & \small\faRobot\normalsize & \small\faRobot\normalsize & \small\faRobot\normalsize 
                      & \footnotesize $+$ & \footnotesize $++$ & \footnotesize Med \\
        \bottomrule
    \end{tabular}
\end{table}

\section{Introduction}

Offline evaluations of information retrieval systems in the Cranfield paradigm~\cite{DBLP:conf/sigir/Cleverdon91} require relevance judgments. However, assessing if a document is relevant to an information need is a highly subjective task~\cite{taube:1965}, which can make experimental observations unreliable~\cite{DBLP:books/sp/19/Voorhees19}. Consequently, information needs are, at least in the traditional setup of the Text Retrieval Conference (TREC)~\cite{10.5555/1121636}, formalized to increase the agreement of relevance assessors. The standard formalization of information needs in so-called topics consists of a title (submitted to the retrieval system), a description (what the user has in mind), and a narrative (often providing criteria for which documents are relevant and which are not). Each of those fields formalizes the information needs at different levels, ensuring that information needs where searchers can not properly formulate queries~(e.g., anomalous states of knowledge~\cite{belkin1980anomalous}) can be evaluated.

Leveraging instruction-tuned LLMs to generate relevance judgments has the potential to substantially reduce the cost of constructing test collections, provided that corpora constructed with LLMs still yield reliable evaluations and observations. Motivated by this possibility, earlier work has examined the quality of LLM-based relevance judgments in ad hoc information retrieval experiments~\cite{faggioliPerspectivesLargeLanguage2023,DBLP:conf/sigir/0001SC024,DBLP:conf/sigir/ArabzadehC25a}, finding that LLM relevance judgments agree to a considerable extent with human assessors and also that the resulting system rankings are similar. Still, it is an ongoing discussion if LLMs should be used for relevance judgments~\cite{DBLP:journals/irrj/Soboroff25}. A frequent challenge in LLM relevance assessment is positivity bias, where models overestimate relevance compared to human annotators~\cite{DBLP:conf/sigir/TakehiVSS25,DBLP:conf/sigir/BalogM025,DBLP:conf/sigir/FrobePSM0PH25}. However, there are scenarios in which LLMs can assist human assessors in (semi-)automated workflows,  making the construction of relevance judgments more efficient and cost-effective~\cite{DBLP:conf/ictir/FarziD25,DBLP:conf/sigir/FarziD25}.

For the relevance assessment process, topics are the key component as they help reduce ambiguity, ensuring that relevance judgments are consistent and accurately reflect the user's intent. For system developers, topics provide the needed information to interpret the query and allow deeper failure analysis. The level of detail provided by topics helps diagnose system weaknesses and drive meaningful improvements in retrieval effectiveness. However, many modern datasets, such as MS~MARCO~\cite{nguyenMSMARCOHuman2016}, are mined from query logs and therefore do not have traditional TREC-style topics. Consequently, it is unclear which information needs the searchers actually had~\cite{sandersonBESTPRACTICESTEST2010}. Importantly, removing the traditional formalization of topics from the relevance assessments might substantially decrease the reliability of the judgments when LLMs do the assessments, as LLMs then have more room to interpret the information need and no incentive to have a consistent judgment behaviour.

Contrary to prior work that studied the quality of LLM relevance judgments without modifying the available formalization of an information need~\cite{faggioliPerspectivesLargeLanguage2023,DBLP:conf/sigir/0001SC024,DBLP:journals/corr/abs-2411-08275}, we specifically ask if LLM assessors should formalize information needs in situations when no TREC-style topic is available. This formalization step only increases the costs slightly (e.g., because the LLM-prompts get longer as the formalization must be injected into the judgment process), but makes the judgment process more similar to the traditional standard of TREC. To this end, we pay special attention to the representations of the information needs that are provided to the LLM, and we formalize the information need into a retrieval topic using information from the search contexts.
Therefore, we formalize topics using LLMs, prompted with search contexts such as queries, query variants, relevant and non-relevant documents, and test seven prompts across different prompting paradigms.
We then use the resulting formalizations with LLM relevance assessors to judge document relevance. We evaluate our approach on the \acl{R04} and the 2019 and 2020 editions of TREC Deep Learning and four open-source LLMs.
Our work is the first to validate whether synthetically formalized topics improve LLM relevance judgments.

Table~\ref{tab:comparison} provides an overview of our approach in comparison to the traditional TREC setting and the default LLM behaviour that does not formalize topics explicitly (e.g., in UMBRELA~\cite{DBLP:journals/corr/abs-2406-06519}). The traditional TREC setup has a long history of using and publishing formalized information needs by humans, which might have contributed to the high reliability (but also high costs) of the resulting corpora. The emerging LLM relevance assessments rely most often only on a query (as many new scenarios do not come with formalized topics) to represent the information need and assess the relevance fully automatically~\cite{DBLP:journals/corr/abs-2406-06519}. This process is much cheaper but less reliable, as recent studies show~\cite{DBLP:conf/sigir-ap/Alaofi0SS24,DBLP:conf/sigir/ArabzadehC25a,DBLP:conf/sigir/FrobePSM0PH25}. In contrast, our approach synthesizes information needs from queries or a limited set of human annotations to construct structured topics for LLM relevance assessment. Our findings show that this additional step increases the agreement of LLMs and helps them judge fewer documents as relevant (mitigating the positivity bias), which thereby improves the reliability. Our main contributions are:

\begin{itemize}
	\item A detailed analysis of LLM relevance judgments from synthetically formalized topics,
	\item an investigation of how these judgments influence IR benchmarks and the reliability of test collections,
	\item a comparative analysis of the similarity between TREC topics and generated counterparts,
	\item and a large dataset containing over 50 thousand generated topics, synthesized from queries and documents, and 1.7 million LLM relevance judgments.
\end{itemize}

All experiments, synthetically formalized topics, code, LLM relevance judgments, and further results are made publicly available.\footnote{https://github.com/irgroup/sigir26-synthesized-information-needs}

\section{Related Work}

System-oriented IR evaluations require reliable resources that are conventionally defined by three components: \Ni a document collection, \Nii human-curated topics that describe information needs, and \Niii corresponding relevance judgments that associate documents with information needs at varying levels of relevance \cite{sandersonBESTPRACTICESTEST2010}. TREC established the de facto standard syntax for representing information needs in systematic IR evaluations \cite{10.5555/1121636}. Although the topic syntax underwent several revisions in early iterations, it has conventionally comprised three elements: a title, a description, and a narrative. This topic structure has been used since the inauguration of TREC in 1992 and was \emph{``designed to mimic a real user's need''}~\cite{DBLP:conf/sigir/Harman93}.

Integrating the concept of LLM-as-a-judge~\cite{DBLP:conf/nips/ZhengC00WZL0LXZ23} into system-oriented IR evaluation has the potential to fully automate systematic assessments of IR systems~\cite{DBLP:conf/sigir/RahmaniCY0C24,DBLP:conf/www/Rahmani0YC0025}. At the core of the retrieval problem lies the question of whether a document is relevant to a given information need. If an LLM can reliably assess relevance between a topic and a document, it could potentially replace human assessors in the evaluation process. This would enable fully automated IR evaluations, provided that both topics and relevance judgments can be synthesized by LLMs. Prior studies have investigated the use of instruction-tuned LLMs to generate synthetic relevance judgments as substitutes for human assessments~\cite{faggioliPerspectivesLargeLanguage2023,DBLP:conf/sigir/0001SC024,DBLP:conf/sigir/TakehiVSS25,DBLP:conf/sigir/FrobePSM0PH25,DBLP:journals/irrj/Soboroff25}.

Seminal works by Faggioli et al.~\cite{faggioliPerspectivesLargeLanguage2023} and Thomas et al.~\cite{DBLP:conf/sigir/0001SC024} report substantial agreement between human and LLM-based relevance judgments. Central to both studies is the use of instruction-tuned LLMs prompted with contextual information, including either a query or a full topic description, along with the documents or passages to be assessed for relevance. Notably, these works evaluate two types of data collections: those in which the information need is represented by short web search queries (MS MARCO as part of TREC Deep Learning~\cite{DBLP:conf/sigir/CraswellMYCVS21}) and those employing more verbose topics that include a title, description, and narrative to articulate the information need (TREC Disks~4\&5 as part of TREC Robust~\cite{DBLP:conf/trec/Voorhees04b}).

Arabzadeh and Culpepper \cite{DBLP:conf/sigir/ArabzadehC25a} present a systematic comparison of multiple relevance judgment generation approaches by evaluating the resulting labels along two complementary dimensions: alignment with human judgments and agreement with system rankings. Human alignment is assessed by discretizing graded relevance labels into three bins and computing agreement via pairwise comparisons between LLM-generated and human judgments. This design is both sensitive to graded relevance and more robust to class imbalance than traditional agreement measures such as Cohen's $\kappa$. Agreement with system rankings is evaluated using the compatibility metric proposed by Clarke et al. \cite{DBLP:conf/ictir/ClarkeVS20}, in which ideal rankings are generated by exhaustively permuting documents with identical relevance labels, and reporting the highest Rank-Biased Overlap (RBO) \cite{DBLP:journals/tois/WebberMZ10} between LLM-induced rankings and these ideal rankings.

Several follow-up studies have identified limitations of LLM-based relevance judgments, including positivity bias \cite{DBLP:conf/sigir/TakehiVSS25,DBLP:conf/sigir/BalogM025}, which can show as score inflation in evaluations, and susceptibility to manipulation, whereby injecting query terms into the evaluated document can nudge the LLM toward a positive relevance decision \cite{DBLP:conf/sigir-ap/Alaofi0SS24}. Fröbe et al. \cite{DBLP:conf/sigir/FrobePSM0PH25} report a high level of agreement among LLMs, which may result in a biased representation of relevance. Similarly, concerns have been raised that synthetic, LLM-based judgments may never achieve the reliability and authenticity of machine-independent, human-generated labels \cite{DBLP:journals/irrj/Soboroff25}. While these concerns are well-founded, a detailed discussion of them is beyond the scope of this work. We acknowledge these limitations and analyze the capabilities of generative LLMs within these constraints.

Earlier work has generally assumed that queries and topic files are given and has focused primarily on automating relevance judgments. Although different modalities of information needs were implicitly considered, few studies systematically analyzed the impact of varying topic representations and configurations on the quality of generated relevance judgments. In this work, we therefore conduct a systematic analysis of how different topic representations affect the quality of LLM-generated relevance judgments.

Closely related to the idea of topic generation is a substantial body of research on query generation \cite{DBLP:conf/sigir/AlaofiGSS023,DBLP:conf/trec/CraswellMYRCLVS23,DBLP:conf/sigir/0002Z0QLJWB24,DBLP:conf/ecir/HosseiniKKVCP25}. Synthetic queries have been employed for a variety of purposes, including query expansion \cite{DBLP:conf/sigir/0002Z0QLJWB24}, document expansion to improve retrieval effectiveness \cite{DBLP:journals/corr/abs-1904-08375}, query performance prediction \cite{DBLP:journals/tois/MengAAAR25}, the simulation of interactive IR experiments \cite{DBLP:conf/ecir/EngelmannBFSF24}, and the diversification of document pools during the construction of test collections \cite{DBLP:conf/cikm/MoffatSTB15}. Like topics, queries represent an information need, specifically, how users articulate their understanding of that need when interacting with an information system.

Several studies suggest that generated queries more accurately represent the underlying information need when additional contextual information is provided \cite{DBLP:conf/sigir/AlaofiGSS023,DBLP:conf/ictir/Alaofi00SS25}. Analogous to query expansion, where supplementary context is introduced through additional query terms, we hypothesize that supplying richer contextual information in the form of more expressive and carefully crafted topic representations can improve the quality of LLM-generated relevance judgments. For instance, Hosseini et al. \cite{DBLP:conf/ecir/HosseiniKKVCP25} propose annotation guidelines for generating queries in the context of product retrieval, in which logged user queries are enriched with extracted requirements and supplementary contextual information. This approach is comparable to the description component of a TREC topic, which may include inclusion and exclusion criteria. Their results indicate that providing additional contextual information improves agreement between human and LLM judgments and enhances the interpretability of the resulting judgments.

We acknowledge that topic and query generation are closely related tasks. If a generated topic is subsequently used to derive a query in a typical ad hoc IR experiment, there is little practical distinction between generating a topic and a query. However, the focus of this work is fundamentally different. We investigate the generation of topics as representations of information needs that are subsequently used to guide an instruction-tuned LLM in producing relevance judgments. To the best of our knowledge, this is the first study to systematically analyze the impact of different topic representations on LLM-generated relevance judgments.

\section{Problem Statement}\label{sec:problem}
In our experimental setting, we posit that a latent information need $I$ is realized as a retrieval topic $T$. Since the information need is an abstract cognitive state rather than an observable variable, the mapping is not a deterministic function; rather, $T$ is a realization drawn from a conditional distribution defined by the user's intent:

\begin{equation}
	T \sim P(T \mid I) \tag{$A_1$} \,. \label{eq:assumption1}
\end{equation}

While this linguistic realization is typically performed by the holder of the cognitive state, in the construction of test collections, this process may be emulated by TREC assessors or organizers who formalize a proxy intent into a title, description, and narrative.

The relevance of a document $d$ from corpus $\mathcal{D}$ to the topic $T$ is assessed by an oracle judge $O$, yielding a set of relevance judgments $\mathcal{R}(T, \mathcal{D})$, commonly referred to as query relevance (qrels). We further assume that the set of judged documents constitutes a sufficient extrinsic representation of the topic. Thus, this set acts as the operational proxy for the topic:
\begin{equation}
	T \cong \mathcal{R}(T, \mathcal{D}) \quad \text{where} \quad \mathcal{R}(T, \mathcal{D}) = \{(d, y) \mid d \in \mathcal{D}, y = O(d, T)\} \,. \tag{$A_2$} \label{eq:assumption2}
\end{equation}

This congruence implies that within the evaluation space, the topic is characterized by its pooled and judged documents. Therefore, if two topics yield similar judgment sets, they are treated as operationally similar:

\begin{equation}
	\mathcal{R}(T_1, \mathcal{D}) \approx \mathcal{R}(T_2, \mathcal{D}) \implies T_1 \approx T_2 \,.
\end{equation}

Consequently, our objective is to identify a generator function $\mathcal{G}$ capable of producing a high-quality topic $\hat{T}$ whose induced relevance judgments approximate the ground truth judgments of the original topic:

\begin{equation}
	\mathcal{R}(\hat{T}, \mathcal{D}) \approx \mathcal{R}(T, \mathcal{D}) \quad \text{where} \quad \hat{T} = \mathcal{G}(\cdot) \,.
\end{equation}

Given these assumptions, we raise the question: \textit{How well can LLMs generate retrieval topics that are useful for producing synthetic relevance judgments?}

\section{Synthesizing Formalized Information Needs}\label{sec:method}

Queries logged from production systems can substitute formal topics to a limited extent, since they reveal only a limited view of the real user's information need. Formal topics include richer details and explicit relevance criteria. Manually creating such topics does not scale, and meeting the growing demand for training data for deep neural IR systems is challenging. Thus, there is great potential in synthesizing structured topics derived from logged queries and their search context.

We approached the synthesis of topics as a text generation task. LLMs are prompted to generate retrieval topics containing a title, description, and narrative. The information need that should be synthesized is represented by contexts from the search, such as queries ($q$), relevant documents ($d^+$), and non-relevant documents ($d^-$), embedded in different prompts. We prioritized these contexts as they are natural to the search setting and often available with minimal manual work. However, the context selection depends highly on the search task and environment.

We explored seven prompts that combine queries, relevant and non-relevant documents in different variations. The general form of the prompt is listed in Figure~\ref{fig:prompt-trecrobust}. While the structure remains the same, the embedded contexts introduce variations, as detailed in Table~\ref{tab:prompts}. The \pcquery{} prompt relies solely on the original query and potential query variants. This prompt is the natural extension of scenarios in which only a query serves as a representation of the information need~\cite{nguyenMSMARCOHuman2016}. This prompt is further extended with relevant or non-relevant documents in the \pcquerydocpos{} and \pcquerydocneg{} prompts. Representing the information need through additional documents requires minimal assessment effort. In return, relevant documents can provide more information about what a user was searching for, and non-relevant documents can be valuable for refining exclusion criteria. In combination, the \pcquerycontrastive{} prompt uses all three components. Contrastive feedback has already been shown to be useful for user simulations with LLMs~\cite{DBLP:conf/sigir/Kruff0S25}.
Further, we also evaluated purely document-based variants: \pcdocpos{}, \pcdocneg{}, and \pccontrastive{}. All prompts were instantiated with a single context example and in a few-shot fashion, with up to five items.

\begin{figure}[t]
	{\sffamily\small \setlength{\fboxsep}{4pt}
\setlength{\fboxrule}{0.8pt}
\sethlcolor{lightgray}
\newcommand{\optionalprompt}[1]{\hl{#1}}
\fbox{%
	\begin{tabular}{p{0.01\columnwidth}p{0.84\columnwidth}}
		                                                                                                                                                            & Given some user \hexrule[3274A1]{.2cm}{.2cm}\hexrule[E1812C]{.2cm}{.2cm}\hexrule[3A9239]{.2cm}{.2cm}\hexrule[C03E3E]{.2cm}{.2cm} \optionalprompt{queries},\hexrule[E1812C]{.2cm}{.2cm}\hexrule[3A9239]{.2cm}{.2cm}\hexrule[9372B2]{.2cm}{.2cm}\hexrule[845B53]{.2cm}{.2cm} \optionalprompt{relevant and} \hexrule[E1812C]{.2cm}{.2cm}\hexrule[C03E3E]{.2cm}{.2cm}\hexrule[9372B2]{.2cm}{.2cm}\hexrule[D684BD]{.2cm}{.2cm}  \optionalprompt{not relevant documents}, you must provide a TREC-style topic that simulates a nuanced user information need. \\
		\\[-5pt]
		                                                                                                                                                            & The topic must include the following fields: title, description, and narrative. The title should be a brief, 2-4 word label for the topic. The description should clearly summarize the user's information goal in one sentence or a question. The narrative should be a short statement that explains the user's intent and the criteria for relevance.                                                                                                                                                                                                \\
		\\[-5pt]
		\multirow{3}{*}{\rotatebox[origin=c]{90}{\hexrule[3274A1]{.2cm}{.2cm}\hexrule[E1812C]{.2cm}{.2cm}\hexrule[3A9239]{.2cm}{.2cm}\hexrule[C03E3E]{.2cm}{.2cm}}} & \optionalprompt{Queries}                                                                                                                                                                                                                                                                                                                                                                                                                                                                                                                                \\
		                                                                                                                                                            & \optionalprompt{A person has typed these queries into a search engine:}                                                                                                                                                                                                                                                                                                                                                                                                                                                                                 \\
		                                                                                                                                                            & \optionalprompt{\textit{<queries>}}                                                                                                                                                                                                                                                                                                                                                                                                                                                                                                                     \\
		\\

		\multirow{3}{*}{\rotatebox[origin=c]{90}{\hexrule[E1812C]{.2cm}{.2cm}\hexrule[3A9239]{.2cm}{.2cm}\hexrule[9372B2]{.2cm}{.2cm}\hexrule[845B53]{.2cm}{.2cm}}} & \optionalprompt{— BEGIN RELEVANT DOCUMENTS CONTENT -}                                                                                                                                                                                                                                                                                                                                                                                                                                                                                                   \\
		                                                                                                                                                            & \optionalprompt{\textit{<relevant\_documents>}}                                                                                                                                                                                                                                                                                                                                                                                                                                                                                                         \\
		                                                                                                                                                            & \optionalprompt{— END RELEVANT DOCUMENTS CONTENT -}                                                                                                                                                                                                                                                                                                                                                                                                                                                                                                     \\
		\\

		\multirow{3}{*}{\rotatebox[origin=c]{90}{\hexrule[E1812C]{.2cm}{.2cm}\hexrule[C03E3E]{.2cm}{.2cm}\hexrule[9372B2]{.2cm}{.2cm}\hexrule[D684BD]{.2cm}{.2cm}}} & \optionalprompt{— BEGIN NOT RELEVANT DOCUMENTS CONTENT -}                                                                                                                                                                                                                                                                                                                                                                                                                                                                                               \\
		                                                                                                                                                            & \optionalprompt{\textit{<not\_relevant\_documents>}}                                                                                                                                                                                                                                                                                                                                                                                                                                                                                                    \\
		                                                                                                                                                            & \optionalprompt{— END NOT RELEVANT DOCUMENTS CONTENT -}                                                                                                                                                                                                                                                                                                                                                                                                                                                                                                 \\
		\\
		                                                                                                                                                            & Output Format and Structure:                                                                                                                                                                                                                                                                                                                                                                                                                                                                                                                            \\
		                                                                                                                                                            & \textit{<format\_instructions>}                                                                                                                                                                                                                                                                                                                                                                                                                                                                                                                         \\

	\end{tabular}
}
}
	\Description{General form of the prompts used to synthesize information needs. \textit{<Italicised>} words are placeholders, filled with appropriate contexts. \colorbox{lightgray}{Shaded} text is optional, included in some prompt variants indicated by the color coding.}
	\caption{General form of the prompts used to synthesize information needs. \textit{<Italicised>} words are placeholders, filled with appropriate contexts. \colorbox{lightgray}{Shaded} text is optional, included in some prompt variants indicated by the color coding.}
	\label{fig:prompt-trecrobust}
\end{figure}

\begin{table}
	\centering
	\caption{Overview of the prompt variants and the contexts (queries ($q$), relevant documents ($d^+$), and non-relevant documents ($d^-$)) they rely on.}
	\small
	\begin{tabular}{llccc}
		\toprule
		                             & \textbf{Prompt}            & \textbf{$q$} & \textbf{$d^+$} & \textbf{$d^-$} \\

		\midrule

		\hexrule[3274A1]{.2cm}{.2cm} & \texttt{query}             & \cmark       & \xmark         & \xmark         \\
		\hexrule[E1812C]{.2cm}{.2cm} & \texttt{query-contrastive} & \cmark       & \cmark         & \cmark         \\
		\hexrule[3A9239]{.2cm}{.2cm} & \texttt{query-docs-pos}    & \cmark       & \cmark         & \xmark         \\
		\hexrule[C03E3E]{.2cm}{.2cm} & \texttt{query-docs-neg}    & \cmark       & \xmark         & \cmark         \\
		\hexrule[9372B2]{.2cm}{.2cm} & \texttt{contrastive}       & \xmark       & \cmark         & \cmark         \\
		\hexrule[845B53]{.2cm}{.2cm} & \texttt{docs-pos}          & \xmark       & \cmark         & \xmark         \\
		\hexrule[D684BD]{.2cm}{.2cm} & \texttt{docs-neg}          & \xmark       & \xmark         & \cmark         \\

		\bottomrule
	\end{tabular}

	\label{tab:prompts}
\end{table}

\section{Evaluation}
What makes a good topic? To the best of our knowledge, no catalog with quality criteria or automated measures exists to directly assess the quality of the formalizations of information needs. This might be because creating topics was an inherently human task, guided by the application area of the test collection.
To assess the viability of synthetic topics for automated relevance judgments, we focus on the quality of labels that can be generated from them.

\subsection{Experimental Setup}

We formalize information needs for three test collections with four LLMs and subsequently use different variants of the formalized information needs to perform relevance judgments with the LLMs.
We evaluate the resulting relevance judgments with respect to the \Ni alignment to human labels and \Nii the agreement of system rankings. Further, we assess \Niii the similarity of the LLM-formalized topics to their human counterparts.

\paragraph{Test Collections}
We synthesized information needs and relevance judgments for Robust04, and the 2019 and 2020 editions of the TREC Deep Learning Tracks (\acs{DL19} and \acs{DL20}).
Robust04 emerged from the \ac{R04}~\cite{DBLP:conf/trec/Voorhees04b} that focused on topics that performed poorly in previous tracks. The track was a traditional ad hoc search task and reused 200 topics from the previous TREC ad hoc tracks from TREC 6–8 and from TREC Robust 2003. Additionally, 50 new topics were added, but for one, no relevant documents were found, bringing the total to 249. Each topic consists of the established fields title, description, and narrative.
The relevance judgments for the old topics were reused from previous TREC evaluations. Participants were allowed to exploit these existing judgments during system development, but it was explicitly forbidden to use them for the run processing~\cite{DBLP:conf/trec/Voorhees04b}.
Pooling based on the old topics showed that about 70.8\,\% of the pooled documents were already judged in the original set. Some runs had unjudged documents at the top ranks because the original judging didn’t cover all newly retrieved documents. For the remaining 49 new topics, a pool of documents was created by selecting the top 100 documents from each participating group's three runs. The assessors who wrote the topics were the same who judged the documents, so the whole topic with all fields was visible, and the narrative contained a concise description of what makes a document relevant~\cite{DBLP:conf/trec/Voorhees04b}.
The assessment of the new topics used a 3-level grading. With this system, the percentage of documents that are labeled relevant is notably lower than for the assessments with binary relevance labels (about 42.1 vs. 76.8).
The 14 participating teams submitted 110 runs, of which 31 runs used only the title field of the topic, and 32 used only the description. As reported, a noticeable difference between the title-only and the description-only runs could be found, and runs that used both fields together were the most effective.
For the LLM relevance judgments, we consider the same stratified sample of 3{,}000 qrels as Thomas et al. before~\cite{DBLP:conf/sigir/0001SC024}. All judgments are made on a scale from 0 (not relevant) to 2 (highly relevant). For binary conversion, the labels 1 and 2 are considered relevant.

The \ac{DL19}~\cite{DBLP:journals/corr/abs-2003-07820} and \ac{DL20}~\cite{DBLP:conf/trec/CraswellMMYC20} test collections are based on the MS MARCO dataset~\cite{nguyenMSMARCOHuman2016}, which contains queries logged from the Bing search engine and sparse relevance labels. While MS MARCO focuses originally on reading comprehension and question answering, selected queries were used for the ad hoc search task.
Since the queries were logged from a live system, no structured topics are available. However, on average, they are almost twice as long as the titles of \ac{R04} and often contain direct questions in natural language. After the pooled judgment process by TREC assessors, the test collections comprise 43 (\ac{DL19}) and 54 (\ac{DL20}) queries.
For both collections, the passage retrieval versions were used as they are established in synthetic judgment experiments. Relevance is judged on a four-label scale from 0 irrelevant, over 1 relevant, 2 highly relevant, up to 3 perfectly relevant. For binary labels, only the two highest grades are considered relevant.

\paragraph{Large Language Models}

\begin{table}
	\caption{Reproduction of the LLM relevance judgments made with GPT-4.1 and the \texttt{-DNA-} prompt by Thomas et al. with open LLMs~\cite{DBLP:conf/sigir/0001SC024}. The label similarity to the TREC queries is measured by Cohen's $\kappa$ and the MAE. The \xmark{} column denotes the number of generation errors. The best results are highlighted in \textbf{bold} and the second best results are \underline{underlined}.}
	\label{tab:agreement_pre_study}
	\resizebox{\columnwidth}{!}{%
		\small
\begin{tabular}{lccc ccc}
\toprule
\textbf{LLM}  & \textbf{Cohen's $\kappa$} & \textbf{MAE} & \textbf{0} & \textbf{1} & \textbf{\xmark}\\
\midrule
GPT-4.1        & .64 \textcolor{gray}{$\pm$ .03} & .17 \textcolor{gray}{$\pm$ .01} & .43 & .57 & -     \\
\midrule
Llama3.1-8B    & .24 \textcolor{gray}{$\pm$ .03} & .42 \textcolor{gray}{$\pm$ .01} & .67 & .33 & 1515  \\
Qwen3-14B      & .70 \textcolor{gray}{$\pm$ .02} & .14 \textcolor{gray}{$\pm$ .01} & .40 & .60 & 32    \\
Mistral3-14B   & .63 \textcolor{gray}{$\pm$ .03} & .18 \textcolor{gray}{$\pm$ .01} & .44 & .56 & 43    \\
gpt-oss-20B    & .64 \textcolor{gray}{$\pm$ .01} & .18 \textcolor{gray}{$\pm$ .01} & .48 & .52 & 8     \\
Qwen3-30B      & .75 \textcolor{gray}{$\pm$ .02} & \underline{.11} \textcolor{gray}{$\pm$ .01} & .29 & .71 & 21    \\
NemoTron3-30B  & .71 \textcolor{gray}{$\pm$ .03} & .13 \textcolor{gray}{$\pm$ .01} & .39 & .61 & 0     \\
Llama3.1-70B   & .76 \textcolor{gray}{$\pm$ .02} & \underline{.11} \textcolor{gray}{$\pm$ .01} & .34 & .66 & 18    \\
Llama3.3-70B   & \textbf{.80} \textcolor{gray}{$\pm$ .02} & \textbf{.09} \textcolor{gray}{$\pm$ .01} & .30 & .70 & 9     \\
Qwen3-Next     & \underline{.77} \textcolor{gray}{$\pm$ .02} & \underline{.11} \textcolor{gray}{$\pm$ .01} & .35 & .65 & 0     \\
gpt-oss-120B   & .72 \textcolor{gray}{$\pm$ .02} & .13 \textcolor{gray}{$\pm$ .01} & .43 & .57 & 1     \\
Deepseek-V3.2  & .71 \textcolor{gray}{$\pm$ .02} & .14 \textcolor{gray}{$\pm$ .01} & .42 & .58 & 20    \\
\bottomrule
\end{tabular}
	}
\end{table}

In a pre-study, we reproduced the \ac{R04} LLM relevance judgments of Thomas et al. with a large variety of open LLMs~\cite{DBLP:conf/sigir/0001SC024}. The results are displayed in Table~\ref{tab:agreement_pre_study}. All LLMs except two match or exceed the performance of GPT-4.1 as used in the original study. The output format was directly included in the prompt, and no guided decoding was used, explaining the many generation errors, especially for smaller models. For this study, we selected \gptlarge{}, \gptsmall{}, \qwenlarge{}, and \qwensmall{}.\footnote{Hugging Face: \gptlarge{}: \href{https://huggingface.co/openai/gpt-oss-120b}{\scriptsize\color{softblue}\faLink}; \gptsmall{}: \href{https://huggingface.co/openai/gpt-oss-20b}{\scriptsize\color{softblue}\faLink}; \qwenlarge{}: \href{https://huggingface.co/Qwen/Qwen3-Next-80B-A3B-Instruct-FP8}{\scriptsize\color{softblue}\faLink}; \qwensmall{}: \href{https://huggingface.co/Qwen/Qwen3-30B-A3B-Instruct-2507}{\scriptsize\color{softblue}\faLink}}
The selected LLMs have a Mixture-of-Experts (MoE) architecture, which means that they use fewer parameters per token and are more efficient compared to dense architectures such as Llama3. This LLM selection allows comparisons across model sizes (small vs. large) and families (GPT vs. Qwen).
All LLMs were used with default configurations, which include a medium reasoning effort for the GPT models. For the Qwen3 models, thinking was disabled.

\begin{description}
	\item[\gptlarge{}] A high-capacity open-weight model released by OpenAI. It utilizes a sparse MoE architecture with 128 experts and a total parameter count of $117B$, of which only $5.1B$ parameters are active per token. This model is optimized for agentic reasoning and complex instruction following~\cite{DBLP:journals/corr/abs-2508-10925}.
	\item[\gptsmall{}] The distilled, medium-scale variant of the gpt-oss family. It comprises 32 experts and $21B$ total parameters with $3.6B$ active parameters per token~\cite{DBLP:journals/corr/abs-2508-10925}. This smaller model is included to explore a speed and cost trade-off for the topic synthesis and automated relevance assessment task.
	\item[Qwen3-Next-80B-A3B (\qwenlarge{})] A next-generation iteration of the Qwen3 series that introduces a highly sparse MoE architecture. It features 512 total experts, activating only $3B$ parameters per token. We use the fine-grained fp8 quantized version for efficiency and included the model as an alternative to \gptlarge{}.
	\item[Qwen3-30B-A3B (\qwensmall{})] A smaller MoE model within the Qwen 3 family featuring $30.5B$ total parameters and $3.3B$ active parameters. Unlike the Next variant, it employs a more traditional MoE structure with 128 experts~\cite{DBLP:journals/corr/abs-2505-09388}. This model is included as a smaller counterpart to \qwenlarge{}.
\end{description}

\subsection{Alignment of Relevance Judgments}

\begin{table*}[t]
	\caption{Label alignment for the \gptlarge{} LLM relevance assessments using synthesized information needs in comparison to the TREC judgments. The alignment is measured by Cohen's $\kappa$ ($\kappa$) and the MAE. A 95\% confidence interval based on 20 bootstraps is reported in light grey. Additionally, the label distribution (0, 1, 2, and 3) is reported. The best results are highlighted in \textbf{bold} and the second best are \underline{underlined}.}
	\label{tab:label_agreement_example}
	\resizebox{\textwidth}{!}{
		\begin{tabular}{llll @{\hskip 15pt} ccccc @{\hskip 15pt}cccccc@{\hskip 15pt} cccccc}
\toprule
\textbf{Prompt} &  \multicolumn{3}{c@{\hskip 15pt}}{\textbf{Context}} & \multicolumn{5}{c@{\hskip 15pt}}{\textbf{\ac{R04}}} & \multicolumn{6}{c@{\hskip 15pt}}{\textbf{\ac{DL19}}} & \multicolumn{6}{c}{\textbf{\ac{DL20}}} \\
 \cmidrule(r{10pt}){2-4} \cmidrule(r{10pt}){5-9} \cmidrule(r{10pt}){10-15} \cmidrule(l){16-21} 
 
 & $q$ & $d^+$ & $d^-$ &  $\kappa$ & MAE & 0 & 1 & 2 & $\kappa$ & MAE & 0 & 1 & 2 & 3 & $\kappa$ & MAE & 0 & 1 & 2 & 3 \\
\midrule

TREC Topic (Baseline) &  &  &  & \textbf{.56}{\textcolor{gray}{±.02}} & \textbf{.40}{\textcolor{gray}{±.01}} & .43 & .33 & .24 & .41{\textcolor{gray}{±.01}} & .67{\textcolor{gray}{±.02}} & .31 & .40 & .13 & .16 & .39{\textcolor{gray}{±.01}} & .61{\textcolor{gray}{±.01}} & .39 & .39 & .09 & .12 \\
\midrule

\pcquery{} & 1 &  &  & .31{\textcolor{gray}{±.02}} & .61{\textcolor{gray}{±.02}} & .60 & .36 & .03 & .37{\textcolor{gray}{±.01}} & .58{\textcolor{gray}{±.01}} & .40 & .49 & .08 & .03 & .40{\textcolor{gray}{±.01}} & .48{\textcolor{gray}{±.01}} & .48 & .42 & .08 & .03 \\

\midrule

\multirow[c]{5}{*}{\pcquerycontrastive{}} & 1 & 1 & 1 & .42{\textcolor{gray}{±.02}} & .51{\textcolor{gray}{±.02}} & .53 & .36 & .12 & \textbf{.46}{\textcolor{gray}{±.01}} & .\textbf{54}{\textcolor{gray}{±.01}} & .38 & .45 & .11 & .07 & \underline{.45}{\textcolor{gray}{±.01}} & .47{\textcolor{gray}{±.01}} & .46 & .39 & .10 & .05 \\
 & 1 & 2 & 2 & .43{\textcolor{gray}{±.02}} & .51{\textcolor{gray}{±.02}} & .51 & .36 & .13 & \underline{.45}{\textcolor{gray}{±.01}} & .\textbf{54}{\textcolor{gray}{±.01}} & .39 & .43 & .13 & .05 & .\underline{45}{\textcolor{gray}{±.01}} & .47{\textcolor{gray}{±.01}} & .48 & .38 & .10 & .05 \\
 & 1 & 3 & 3 & \underline{.44}{\textcolor{gray}{±.02}} & \underline{.49}{\textcolor{gray}{±.02}} & .50 & .36 & .14 & \underline{.45}{\textcolor{gray}{±.01}} & .\textbf{54}{\textcolor{gray}{±.01}} & .38 & .45 & .12 & .05 & .\underline{45}{\textcolor{gray}{±.01}} & .46{\textcolor{gray}{±.01}} & .48 & .40 & .09 & .04 \\
 & 1 & 4 & 4 & \underline{.44}{\textcolor{gray}{±.02}} & \underline{.49}{\textcolor{gray}{±.01}} & .51 & .36 & .13 & \textbf{.46}{\textcolor{gray}{±.02}} & .\textbf{54}{\textcolor{gray}{±.01}} & .38 & .43 & .13 & .06 & .\underline{45}{\textcolor{gray}{±.01}} & .\underline{46}{\textcolor{gray}{±.01}} & .48 & .38 & .10 & .05 \\
 & 1 & 5 & 5 & \underline{.44}{\textcolor{gray}{±.02}} & \underline{.49}{\textcolor{gray}{±.02}} & .51 & .36 & .13 & .44{\textcolor{gray}{±.01}} & .\textbf{54}{\textcolor{gray}{±.01}} & .39 & .45 & .12 & .05 & \textbf{.46}{\textcolor{gray}{±.01}} & .\textbf{45}{\textcolor{gray}{±.01}} & .48 & .39 & .10 & .04 \\
\midrule

\multirow[c]{5}{*}{\pcquerydocpos{}} & 1 & 1 &  & .36{\textcolor{gray}{±.02}} & .57{\textcolor{gray}{±.02}} & .57 & .35 & .07 & .41{\textcolor{gray}{±.01}} & .56{\textcolor{gray}{±.01}} & .37 & .50 & .09 & .04 & .43{\textcolor{gray}{±.01}} & .48{\textcolor{gray}{±.01}} & .46 & .42 & .08 & .04 \\
 & 1 & 2 &  & .37{\textcolor{gray}{±.02}} & .56{\textcolor{gray}{±.02}} & .56 & .37 & .07 & .42{\textcolor{gray}{±.01}} & .56{\textcolor{gray}{±.01}} & ..38 & .48 & .10 & .04 & .43{\textcolor{gray}{±.01}} & .48{\textcolor{gray}{±.01}} & .47 & .41 & .09 & .03 \\
 & 1 & 3 &  & .37{\textcolor{gray}{±.02}} & .56{\textcolor{gray}{±.02}} & .55 & .37 & .08 & .42{\textcolor{gray}{±.01}} & \underline{.55}{\textcolor{gray}{±.01}} & .37 & .50 & .10 & .03 & .44{\textcolor{gray}{±.01}} & .46{\textcolor{gray}{±.01}} & .46 & .42 & .09 & .02 \\
 & 1 & 4 &  & .35{\textcolor{gray}{±.02}} & .57{\textcolor{gray}{±.01}} & .56 & .38 & .07 & .43{\textcolor{gray}{±.01}} & \underline{.55}{\textcolor{gray}{±.02}} & .38 & .47 & .10 & .05 & .44{\textcolor{gray}{±.01}} & .45{\textcolor{gray}{±.01}} & .47 & .42 & .09 & .02 \\
 & 1 & 5 &  & .38{\textcolor{gray}{±.02}} & .55{\textcolor{gray}{±.02}} & .54 & .39 & .07 & .41{\textcolor{gray}{±.01}} & .56{\textcolor{gray}{±.01}} & .39 & .50 & .09 & .03 & .45{\textcolor{gray}{±.01}} & .46{\textcolor{gray}{±.01}} & .47 & .41 & .08 & .03 \\

 \midrule
\multirow[c]{5}{*}{\pcquerydocneg{}} & 1 &  & 1 & .33{\textcolor{gray}{±.02}} & .60{\textcolor{gray}{±.02}} & .61 & .33 & .06 & .39{\textcolor{gray}{±.01}} & .58{\textcolor{gray}{±.01}} & .41 & .47 & .08 & .04 & .42{\textcolor{gray}{±.01}} & .47{\textcolor{gray}{±.01}} & .48 & .41 & .09 & .02 \\
 & 1 &  & 2 & .32{\textcolor{gray}{±.02}} & .60{\textcolor{gray}{±.02}} & .61 & .34 & .05 & .40{\textcolor{gray}{±.01}} & .57{\textcolor{gray}{±.01}} & .41 & .48 & .08 & .04 & .40{\textcolor{gray}{±.01}} & .47{\textcolor{gray}{±.01}} &  .48 & .43 & .08 & .01 \\
 & 1 &  & 3 & .33{\textcolor{gray}{±.02}} & .60{\textcolor{gray}{±.02}} & .60 & .34 & .06 & .40{\textcolor{gray}{±.01}} & .57{\textcolor{gray}{±.01}} & .38 & .50 & .08 & .04 & .41{\textcolor{gray}{±.01}} & .48{\textcolor{gray}{±.01}} & .48 & .41 & .08 & .02 \\
 & 1 &  & 4 & .31{\textcolor{gray}{±.02}} & .61{\textcolor{gray}{±.02}} & .61 & .34 & .05 & .40{\textcolor{gray}{±.01}} & .56{\textcolor{gray}{±.01}} & .39 & .49 & .08 & .04 & .42{\textcolor{gray}{±.01}} & .47{\textcolor{gray}{±.01}} & .48 & .41 & .08 & .02 \\
 & 1 &  & 5 & .32{\textcolor{gray}{±.02}} & .62{\textcolor{gray}{±.02}} & .63 & .32 & .06 & .42{\textcolor{gray}{±.01}} & \underline{.55}{\textcolor{gray}{±.01}} & .39 & .49 & .09 & .04 & .42{\textcolor{gray}{±.01}} & .47{\textcolor{gray}{±.01}} & .49 & .41 & .08 & .02 \\

\bottomrule
\end{tabular}

	}
\end{table*}

\begin{figure}[t]
	\centering
	\Description{Comparison of the label alignment between TREC and LLM judgments that rely on synthesized topics. Only prompts that consider the query are considered in this plot.}\includegraphics[width=\linewidth]{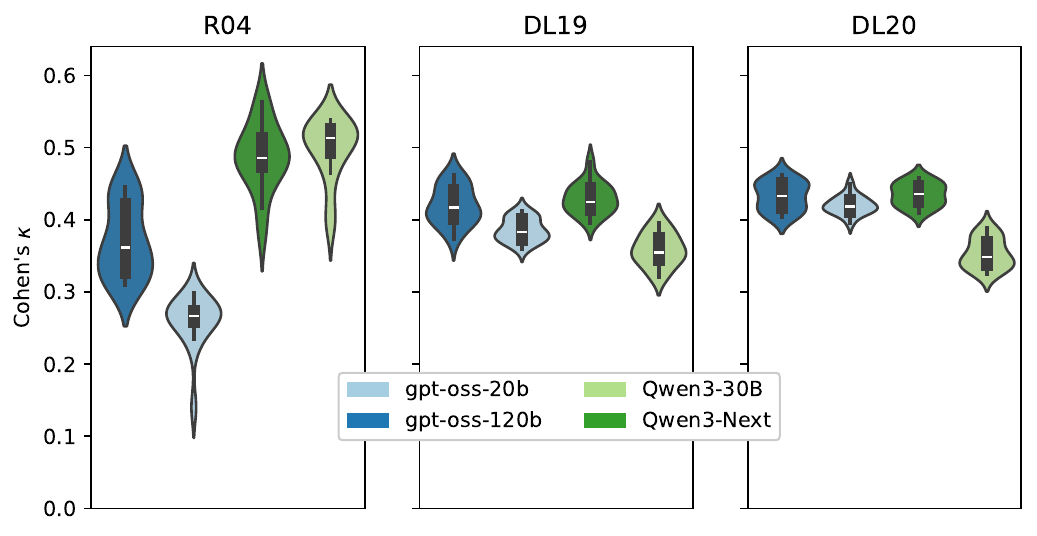}
	\caption{Comparison of the label alignment between TREC and LLM judgments that rely on synthesized topics. Only prompts that consider the query are considered in this plot.}
	\label{fig:label-agreement-llm}
\end{figure}

\begin{figure}[t]
	\centering
	\Description{Label agreement distribution of TREC and LLM judgments that use synthetic topics per context level and across all prompts for \ac{R04}.}        \includegraphics[width=\linewidth]{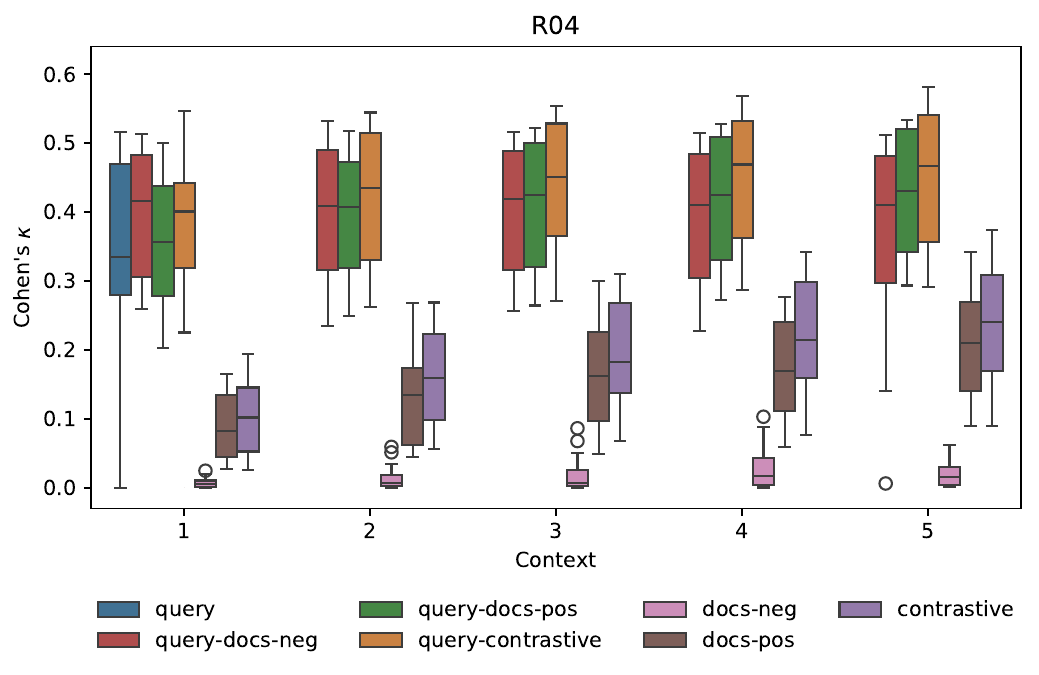}
	\caption{Label agreement distribution of TREC and LLM judgments that use synthetic topics per context level and across all prompts for \ac{R04}.}
	\label{fig:label-agreement-prompt}
\end{figure}

To supplement human annotators, LLM relevance assessors should assign the same relevance label as a human would. If providing a structured topic to an LLM assessor yields similar or improved label alignment, the structured topic can substitute the original representation of the information need. To test this, we compare the synthetic LLM assessments to the human judgments as proposed by Faggioli et al.~\cite{faggioliPerspectivesLargeLanguage2023} and Thomas et al.~\cite{DBLP:conf/sigir/0001SC024}. The document label agreement is measured by Cohen's $\kappa$~\cite{doi:10.1177/001316446002000104} that ranges between 1 for perfect agreement and -1 for complete disagreement. Further, we measure the Mean Average Error (MAE), where 0 denotes perfect agreement and 1 complete disagreement. Following Thomas et al., each similarity measure is reported with a 95 \% confidence interval based on 20 bootstraps across qrels. The bootstraps are also used to measure if the observed differences are significant. Therefore, a two-tailed paired t-test with $\alpha = 0.05$ was conducted across all prompt versions per model and dataset~\cite{DBLP:conf/sigir/0001SC024}. Additionally, we also report the label distribution across relevance grades.

When judging the original TREC topics, the \textbf{open models used in our study outperform GPT-4.1 and GPT-4o most of the time}. Thomas et al. generated judgments for \ac{R04} with GPT-4.1 and achieved a Cohen's $\kappa$ of 0.64 (binary label), which is at least matched by the LLMs we used (see Table~\ref{tab:agreement_pre_study})~\cite{DBLP:conf/sigir/0001SC024}. Upadhyay et al. relied on GPT-4o for \ac{DL19} and \ac{DL20} and achieved a $\kappa$ of 0.36 and 0.35 for the graded judgments, which \gptlarge{} exceeds with 0.41 and 0.39, respectively (see Table~\ref{tab:label_agreement_example})~\cite{DBLP:journals/corr/abs-2406-06519}. These results show that open models can be used for relevance judgments, in line with what Farzi and Dietz observed before~\cite{DBLP:conf/sigir/FarziD25}.
Overall, the similarity for graded relevance judgments is lower compared to the binary judgments, which is consistent with previous observations~\cite{DBLP:journals/corr/abs-2406-06519,DBLP:conf/sigir/FarziD25,faggioliPerspectivesLargeLanguage2023}.

Due to space constraints, we focus on the graded results of \gptlarge{} and the four query-dependent prompts reported in Table~\ref{tab:label_agreement_example}.\footnote{Results of other model sizes and families are only reported in the text.} The results show that \textbf{LLM relevance judgments that use structured information needs can outperform the LLM judgments on the original TREC topics for \ac{DL19} and \ac{DL20}}. For both test collections, the topics consist only of a user query. For \ac{DL19}, the label similarity for LLM judgments that use topics formalized with \pcquerycontrastive{} is significantly better compared to the qrels judged by the same LLM but rely only on the original \pcquery{}. One relevant and one non-relevant document are already sufficient to improve similarity across all LLMs except \gptsmall{} in the graded setting. The best results are achieved with the \pcquerycontrastive{} prompt, but the prompts \pcquerydocpos{} and \pcquerydocneg{} can also improve label similarity.
Similar results are also achieved on \ac{DL20}. The \pcquerycontrastive{} prompt consistently produces higher label similarities and lower MAE. At best, at with five contexts, \gptlarge{} achieves the highest Cohen's $\kappa$ of 0.46 and significantly improves on the qrels generated by using only the TREC topic.

On \ac{R04}, the label similarity based on the original TREC topics that contain all topic fields‚ could not be matched. At best, \gptlarge{} achieves a Cohen's $\kappa$ of 0.44 with the \pcquerycontrastive{} prompt and a context of one query and at least three relevant and not relevant documents. In comparison, using the TREC topics results in a $\kappa$ of over 0.56 for the same model. The Qwen LLMs perform better on this test collection and \qwenlarge{} matches the $\kappa$ of 0.56 achieved with the TREC topics by \gptlarge{} with four context documents \qwenlarge{}. However, this LLM also achieves a higher baseline of 0.6 on the TREC topics (see also Figure \ref{fig:label-agreement-llm}).

Figure~\ref{fig:label-agreement-prompt} shows the $\kappa$ spread across all prompts for \ac{R04}. The differences in $\kappa$ show that~\textbf{the label agreement is highly dependent on the topic prompt}. This highlights the potential of prompt engineering and context selection for topic formalization and relates well to the observation that even small changes in the annotation prompt significantly influence the label quality~\cite{DBLP:conf/sigir/0001SC024}. For example, Cohen's $\kappa$ ranges between almost zero for the \pcdocneg{} prompt and 0.6 for the \pcquerycontrastive{} prompt at a context of five queries, relevant, and non-relevant documents, judged by \qwenlarge{}.
In comparison, Figure~\ref{fig:label-agreement-llm} shows the distribution per LLM across all three test collections for the four prompts that use the query. The LLM has a smaller influence compared by the range between average $\kappa$ scores across LLMs, especially if we compare LLMs per model family.

All prompts were also instantiated in a few-shot-like setting with different context sizes, e.g., the number of documents provided for synthesis. \textbf{Most prompts show the largest improvement up to a context of two or three}. Adding further context gives only diminishing returns. The absolute peak is still most of the time at the highest tested context of five.

Furthermore, we tested the influence of the query as context by synthesizing topics only from relevant and non-relevant documents. The results in Figure \ref{fig:label-agreement-prompt} show a substantially lower label similarity for the prompts without a query. This highlights that \textbf{the original query is an important feature for the topic synthesis.} Topic synthesized only from not relevant documents (\pcdocneg{}) achieve a barely better than random Cohen's $\kappa$ of 0.09 at best. If no query is available, combining relevant and not relevant documents in the \pcontrastive{} prompt achieves the highest similarity, and more context further improves it.
Since the query proved to be an important feature for topic synthesis, we further considered query variants as a context source. Therefore, we provided up to four query variants from the User Query Variants (UQV) dataset, additionally to the original query~\cite{DBLP:conf/adcs/BenhamC17}.
The results on \ac{R04} show that providing more queries to the \pquery{} prompt does not improve similarity or reduce error. When additional query variants are provided to the \pcquerycontrastive{} prompt, the similarity is only sometimes improved by a few percentage points.

\begin{table}[t]
	\centering
	\caption{Complete results of the Multiple Linear Regression models with dummy coding. Coefficients ($\beta$) represent the marginal influence on Cohen's $\kappa$ agreement. The baseline (Intercept) represents the \gptlarge{} for the judge LLM and \qwensmall{} for the topic LLM, and the \pcquery{} prompt.}
	\label{tab:ols}

	\resizebox{\columnwidth}{!}{
		\begin{tabular}{lllll}
			\toprule
			\textbf{}                                                                     & \textbf{Variable}     & \textbf{\ac{DL19} ($\beta$)} & \textbf{\ac{DL20} ($\beta$)} & \textbf{\ac{R04} ($\beta$)} \\ \midrule

			\textbf{Baseline}                                                             & Intercept             & +0.410$^{***}$               & +0.422$^{***}$               & +0.299$^{***}$              \\ \addlinespace
			\textbf{Judge}                                                                & \gptsmall{}           & -0.023$^{***}$               & -0.008$^{***}$               & -0.073$^{***}$              \\
			                                                                              & \qwenlarge{}          & -0.003                       & -0.002                       & +0.119$^{***}$              \\
			                                                                              & \qwensmall{}          & -0.065$^{***}$               & -0.075$^{***}$               & +0.153$^{***}$              \\ \addlinespace
			\textbf{Topic}                                                                & \gptsmall{}           & -0.020$^{***}$               & -0.005$^{**}$                & -0.006                      \\
			                                                                              & \gptlarge{}           & -0.008$^{*}$                 & +0.001                       & +0.030$^{***}$              \\
			                                                                              & \qwenlarge{}          & +0.006                       & +0.002                       & +0.022$^{**}$               \\ \addlinespace

			\textbf{Prompt}                                                               & \pcquerydocneg{}      & +0.007                       & -0.001                       & +0.013                      \\
			                                                                              & \pcquerydocpos{}      & +0.017$^{**}$                & +0.012$^{***}$               & +0.020                      \\
			                                                                              & \pcquerycontrastive{} & +0.032$^{***}$               & +0.017$^{***}$               & +0.028                      \\ \addlinespace

			\textbf{Affinity}                                                             & Same LLM              & -0.003                       & -0.007$^{***}$               & -0.001                      \\

			\textbf{Context}                                                              & \# context docs       & +0.001                       & +0.002$^{***}$               & +0.007$^{***}$              \\ \midrule

			\textbf{Model Fit}                                                            & $R^2$                 & 0.713                        & 0.917                        & 0.823                       \\
			                                                                              & Adj. $R^2$            & 0.700                        & 0.913                        & 0.815                       \\ \bottomrule
			\multicolumn{4}{l}{\small $^{*}$p < 0.05, $^{**}$p < 0.01, $^{***}$p < 0.001} & {\small n = 256}                                                                                                  \\

		\end{tabular}}
\end{table}

To disentangle the factors influencing the alignment, we fit an Ordinary Least Squares (OLS) regression model with dummy coding to estimate the marginal effect ($\beta$) of each system component on Cohen's $\kappa$. The analysis was repeated for each test collection independently, and the results are displayed in Table~\ref{tab:ols}. Our isolated components successfully explained the majority of the variance in $\kappa$ scores, achieving strong adjusted $R^2$ values of 0.7, 0.913, and 0.815, respectively.
To determine which variable had the strongest overall impact, we conducted a Type II ANOVA and calculated the $\eta^2$ effect sizes. It showed that the LLM used for the judgments has a significant influence on the variance in $\kappa$ ($p<0.001$ across all datasets). On the \ac{R04}, for example, the judging LLM accounted for 77.6\% of the total variance in $\kappa$ ($\eta^2 = 0.776, p<0.001$). However, the direction of this influence depends on the domain. For instance, \qwensmall{} judgments do not align well with the \ac{DL19} and \ac{DL20} test collections (coefficients of -0.065 and -0.075) but work well on \ac{R04} (+0.153) (Table~\ref{tab:ols}). Despite these shifts in judge behavior, prompt performance remained stable, with the \pcquerycontrastive{} strategy consistently delivering the highest positive impact. The results yielded no evidence of LLM self-bias. When an LLM judges topics generated by the same LLM, the effect was not significant on \ac{DL19} and \ac{R04}. While it was significant on \ac{DL20}, it actually resulted in a negligible, negative influence (-0.007). Finally, expanding the number of context documents provided a minor but significant improvement in $\kappa$ alignment on both \ac{DL20} and \ac{R04}.

\subsection{Reusability of LLM Relevance Judgments with Formalized Information Needs}

Ideally, retrieval experiments are reliable so that repetitions under similar conditions would yield similar outcomes with high probability~\cite{DBLP:books/sp/19/Voorhees19}. For instance, when relevance judgments are done with different annotators, the outcomes of reliable experiments would still be similar (e.g., when annotators are well-trained and when topics are formalized). Two main aspects can impact reliability: the subjectivity of relevance judgments and the incompleteness of relevance judgments~\cite{DBLP:books/sp/19/Voorhees19}. Measuring agreement among multiple annotators can identify cases in which an annotation task is too subjective for reliable experimentation, and, conversely, highly subjective annotations yield lower inter-annotator agreement. The incompleteness can be measured using leave-one-group-out tests~\cite{DBLP:conf/sigir/Zobel98} that verify if the evaluation is also reliable for retrieval systems that did not contribute to the pool. If there is a serious problem with incompleteness, an evaluation corpus is not reusable to evaluate new systems, whereas a low incompleteness problem indicates that new systems can be evaluated. Furthermore, prior work has shown that too many relevant documents in an evaluation scenario can make evaluation resources non-reusable~\cite{DBLP:conf/sigir/VoorheesCL22} (i.e., leave-one-group-out tests indicated that new systems are not evaluated reliably when there are too many relevant documents). LLMs are known to be much more positive than human relevance assessors~\cite{DBLP:conf/sigir/BalogM025,DBLP:conf/sigir/FrobePSM0PH25}. Hence, the positivity of LLM judgments might reduce the re-usability of corpora with LLM judgments. Well-formalized topics should reduce both threats, subjectivity and incompleteness. Better formalization leaves less room for interpretation, which should increase the agreement among assessors. Similarly, a well-formalized information need should also reduce the number of relevant documents, as the relevance criteria is formalized before the annotations are done.

\begin{table}[t]
\caption{Overview of how different topic formalizations, using only the title (i.e., no formalization), respectively combinations of title, description, and narrative impact the agreement measured as Fleiss Kappa of four LLM relevance assessors respectively the percentage of documents the considered relevant. High agreement respectively lower percentage of relevant documents are better for reliable experimentation.}
\label{tab:agreement_of_topic_features}
\small
\begin{tabular}{@{}ccccccc}
\toprule
\multicolumn{3}{c}{\bf Topic} & \multicolumn{2}{c}{\bf Agreement} & \multicolumn{2}{c}{\bf Relevant} \\

\cmidrule(r){1-3}
\cmidrule(lr){4-5}
\cmidrule(l){6-7}

Title & Desc. & Narr. & DL19 & DL20 & DL19 & DL20 \\

\midrule

\cmark & \xmark & \xmark & 38.85 & 42.95 & 70.67 & 63.44\\
\cmark & \cmark & \xmark & 42.41 & 46.08 & 65.8 & 58.16\\
\cmark & \xmark & \cmark & 44.34 & 46.72 & 60.11 & 51.52\\
\cmark & \cmark & \cmark & 43.22 & 48.53 & 59.81 & 50.98\\

\midrule 

\multicolumn{3}{l}{Original} & --- & --- & 44.30 & 31.70\\

\bottomrule
\end{tabular}
\end{table}

Table~\ref{tab:agreement_of_topic_features} shows the agreement and the percentage of relevant documents for the LLM relevance judgments that use the different topic formalizations. We report the agreement as Fleiss Kappa~\cite{zis-Fleiss1971Measuring}, showing that skipping topic formalization and relying only on the query for relevance judgments yields the lowest inter-annotator agreement. \textbf{Using more formalized topics always increases the agreement}. Completely formalized topics, including a title, description, and narrative, yield higher overall agreement than less formalized variants and no formalization in particular. Skipping the topic formalization also results in substantially more documents being judged as relevant (yielding 70.67\,\% and 63.44\,\% relevant for \ac{DL19} and \ac{DL20}, respectively). \textbf{Using more formalization reduces the number of documents that are judged as relevant, by more than 10\,\% points} for full formalization (with title, description, and narrative). Still, LLM judgments remain more positive than human judgments (the original row reports the portion of documents that are relevant in the human judgments; please note that no agreement can be computed in this scenario). Overall, our formalization improves the agreement of LLM assessors and reduces their positivity, which should yield more reliable experimentation.

\begin{table}
	\caption{Similarity between original topic components of the Robust dataset and components reconstructed by \gptlarge{} for different input configurations.}
	\label{tab:similarity_gen_topics_masked}
	\resizebox{\columnwidth}{!}{%
		\renewcommand{\arraystretch}{1.2}
\setlength{\tabcolsep}{10pt}
\small
\begin{tabular}{llccc}

\toprule
\textbf{Input} & \textbf{Reconstructed} & \textbf{BERTScore} & \textbf{Rel. Len} & \textbf{RougeL} \\
\midrule

Description & Title & \cellcolor{heatmax!100}\color{white}0.72 & \cellcolor{heatmax!0}\color{black}1.21 & \cellcolor{heatmax!100}\color{white}0.58 \\
Narrative & Title & \cellcolor{heatmax!82}\color{white}0.69 & \cellcolor{heatmax!2}\color{black}1.23 & \cellcolor{heatmax!81}\color{white}0.50 \\
Desc. \& Narr. & Title & \cellcolor{heatmax!100}\color{white}0.72 & \cellcolor{heatmax!5}\color{black}1.26 & \cellcolor{heatmax!100}\color{white}0.58 \\

\midrule

Title & Description & \cellcolor{heatmax!70}\color{white}0.67 & \cellcolor{heatmax!19}\color{black}1.41 & \cellcolor{heatmax!31}\color{black}0.28 \\
Narrative & Description & \cellcolor{heatmax!58}\color{black}0.65 & \cellcolor{heatmax!65}\color{white}1.89 & \cellcolor{heatmax!27}\color{black}0.26 \\
Title \& Narr. & Description & \cellcolor{heatmax!70}\color{white}0.67 & \cellcolor{heatmax!65}\color{white}1.89 & \cellcolor{heatmax!36}\color{black}0.30 \\

\midrule

Title & Narrative & \cellcolor{heatmax!0}\color{black}0.55 & \cellcolor{heatmax!100}\color{white}2.25 & \cellcolor{heatmax!0}\color{black}0.14 \\
Description & Narrative & \cellcolor{heatmax!5}\color{black}0.56 & \cellcolor{heatmax!90}\color{white}2.15 & \cellcolor{heatmax!6}\color{black}0.17 \\
Title \& Desc. & Narrative & \cellcolor{heatmax!11}\color{black}0.57 & \cellcolor{heatmax!54}\color{white}1.78 & \cellcolor{heatmax!9}\color{black}0.18 \\

\bottomrule
\end{tabular}
	}
\end{table}

\begin{table*}
\caption{Overview of Leave-one-Group-Out experiments when using different topic formalizations (combinations with respectively without description and narrative) to create the relevance judgments on Deep Learning 2019/2020. We report the Spearman and TauAP ranking correlations (higher is more re-usable) for nDCG@10, nDCG@20, and nDCG@1000.}
\label{tab:table_leave_one_group_out}
\begin{tabular}{@{}ccccccccccccccc}
\toprule
\multicolumn{3}{c}{\bf Topic Formalization} & \multicolumn{6}{c}{\bf Spearman} & \multicolumn{6}{c}{\bf TauAP} \\

\cmidrule(r){1-3}
\cmidrule(lr){4-9}
\cmidrule(l){10-15}

Title & Descr. & Narr. & \multicolumn{2}{c}{\bf nDCG@10} & \multicolumn{2}{c}{\bf nDCG@20} & \multicolumn{2}{c}{\bf nDCG} & \multicolumn{2}{c}{\bf nDCG@10} & \multicolumn{2}{c}{\bf nDCG@20} & \multicolumn{2}{c}{\bf nDCG} \\

\cmidrule(lr){4-5}
\cmidrule(lr){6-7}
\cmidrule(lr){8-9}
\cmidrule(lr){10-11}
\cmidrule(lr){12-13}
\cmidrule(l){14-15}

&&& DL19 & DL20 & DL19 & DL20 & DL19 & DL20 & DL19 & DL20 & DL19 & DL20 & DL19 & DL20\\

\midrule

\cmark & \xmark & \xmark & 0.984 & 0.997 & 0.978 & 0.992 & 0.755 & 0.878 & 0.942 & 0.977 & 0.829 & 0.883 & 0.584 & 0.678\\
\cmark & \cmark & \xmark & 0.988 & 0.998 & 0.983 & 0.994 & 0.763 & 0.901 & 0.943 & 0.975 & 0.844 & 0.909 & 0.619 & 0.714\\
\cmark & \xmark & \cmark & 0.988 & 0.998 & 0.983 & 0.994 & 0.767 & 0.906 & 0.943 & 0.974 & 0.853 & 0.912 & 0.612 & 0.724\\
\cmark & \cmark & \cmark & 0.988 & 0.998 & 0.984 & 0.994 & 0.776 & 0.910 & 0.944 & 0.976 & 0.864 & 0.916 & 0.621 & 0.728\\

%

\bottomrule
\end{tabular}
\end{table*}

To verify if the reduced positivity of our topic formalization also helps to improve the reusability of the relevance judgments (as suggested by Voorhees et al.~\cite{DBLP:conf/sigir/VoorheesCL22}), we run leave-one-group-out~\cite{DBLP:conf/sigir/Zobel98} tests on all relevance judgments that we have created with different degrees of topic formalization. We use all runs that are submitted to the corresponding TREC tracks and rank all systems by their effectiveness according to nDCG@10, nDCG@20, and nDCG@1000, each using every one of our created relevance judgments. We then iterate over each participating group and modify the relevance judgments so that documents contributed only by this team's runs are removed. On this new set of judgments (that is, reduced in size due to the teams left out), we rank all systems again by their effectiveness using the corresponding evaluation measure. Finally, we calculate how similar both system rankings (based on the complete pool vs. the reduced pool without the documents contributed solely by a team) are in terms of the Spearman ranking correlation, and the TauAP~\cite{DBLP:conf/sigir/YilmazAR08,DBLP:conf/ictir/UrbanoM17} ranking correlation. Spearman treats all positions in a system ranking equally, whereas TauAP gives greater weight to the top positions (i.e., changes that appear only at low ranks are less important to TauAP). Table~\ref{tab:table_leave_one_group_out} shows the results of our leave-one-group-out experiments for different topic formalizations. We observe that no topic formalization (i.e., using only the title) yields the lowest correlations for both Spearman and TauAP, especially when more documents are included in the evaluation, as for nDCG@20 or nDCG. Overall, full formalization (with title, description, and narrative) often achieves the highest correlations, indicating that \textbf{properly formalized topics increase the re-usability} of the resulting corpora.

\subsection{Similarity of Synthetic and TREC Topics}

Finally, we directly assess the synthetically formalized topics by comparing them to their human counterparts.
First, we investigate how well an LLM can reconstruct an incomplete topic. Therefore, the LLMs are prompted with one or two fields of an \ac{R04} topic and are asked to generate the remaining topic field(s). The reconstruction quality is assessed as the similarity of the synthesized topic to the reference topic by the BERTScore~\cite{DBLP:conf/iclr/ZhangKWWA20} for semantic similarity and Rouge~\cite{lin-2004-rouge} for lexical similarity.
Rouge measures the longest common sequence between the two texts, providing insights into lexical similarity. BERTScore captures the semantic similarity between topics, which is important for semantic IR systems. Additionally, we assess the length of the generated fields in relation to the original fields.

The reconstruction results in Table~\ref{tab:similarity_gen_topics_masked} show that
\textbf{semantic reconstruction is easier than reconstructing the original vocabulary.}
Overall, BERTScore remains relatively high across all topic fields and LLMs (0.72 to 0.55), while Rouge ranges from 0.58 to 0.14. Reconstructing the title is the easiest, followed by the description and the narratives. For reconstructing the title, the description is the most important input, and additionally, providing the narrative does not improve the similarity. Similarly, it can be observed that the title is more important than the narrative when the description is reconstructed. Reconstructed narratives show the lowest overall similarity and almost no difference across input fields. These results recommend redundancy between the title and description.
The \textbf{reconstructed topic fields are longer} compared to the reference topic fields. \qwenlarge{} is almost matching the lengths of titles, when provided with a single field, but often doubles it for the descriptions. Providing the narrative field as context often results in an increased length. Besides that, providing two topic fields as input does not result in a longer output compared to one input field.
All LLMs perform similarly and can reconstruct the topic in most cases. Only \gptlarge{} could not generate one out of the 248 topics when provided with a single component.

\begin{figure*}[t]
	\Description{BERTScore similarity between topics synthesized by \gptlarge{} and the \ac{R04} topics by prompt, compared across the whole topic and for single components.}
	\centering
	\begin{minipage}{0.66\textwidth}
		\centering
		\includegraphics[width=\linewidth]{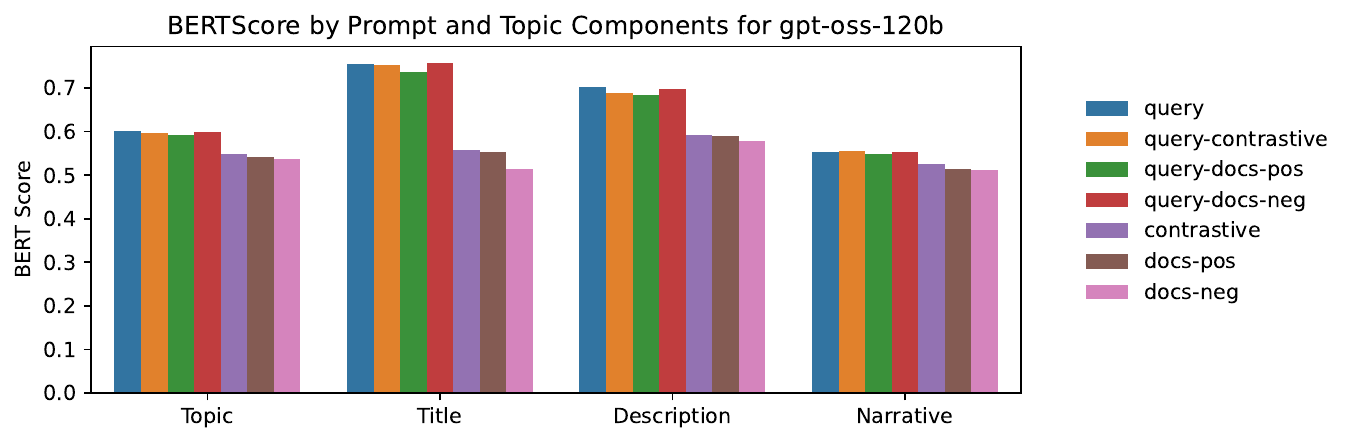}
		\caption{BERTScore similarity between topics synthesized by \gptlarge{} and the \ac{R04} topics by prompt, compared across the whole topic and for single components.}
		\label{fig:topic-similarity-bar}
	\end{minipage}
	\hfill
	\begin{minipage}{0.3\textwidth}
		\centering
		\includegraphics[width=\linewidth]{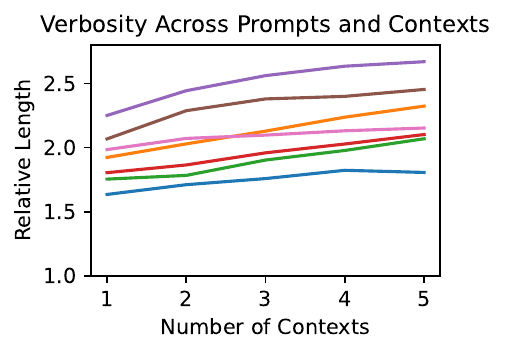}
		\caption{Relative length of the topics synthesized by \gptlarge{} compared to the \ac{R04} reference topics.}
		\label{fig:topic-similarity-len}
	\end{minipage}
\end{figure*}

We further compare the newly synthesized topics to the original topics of \ac{R04} in the same way. Figure~\ref{fig:topic-similarity-bar} shows the BERTScore of the topics synthesized with \gptlarge{} at a context of one.
Overall, only minor differences in similarity can be observed. Across all results, the BERTScore ranges between 0.61 (\pcquerycontrastive{}) and 0.53. (\pcdocneg{}) and Rouge between 0.24 (\pcquery{} and \pcquerycontrastive{}) and 0.1 (\pcdocneg{}). The prompts that use the query perform better than those that do not. However, the differences in these groups are low.

Similar to the reconstruction results, the similarity of the synthetic topics is greater for the title and description fields than for the narrative fields. While the differences are small, different prompts achieve the best results across components. The most similar titles are synthesized by the \pcquery{}. The best descriptions are synthesized with the \pcquery{} prompt, and \textbf{narratives often profit from relevant and not relevant documents} as provided with the \pcquerycontrastive{} prompt, especially for the larger models.

As observed before, the synthetic topics are longer, and adding more context further increases the length, as displayed in Figure~\ref{fig:topic-similarity-len}. While with the right prompt and LLM, synthetic titles can almost match the length of the reference title, the longer fields and especially the narrative are also synthesized longer. The results also show that the Qwen models are more verbose than the GPT models.

In summary, a single query is enough to synthesize a topic title and description with high similarity, and narratives are best synthesized from queries and relevant and non-relevant documents. It is noteworthy that we were able to synthesize topic titles, descriptions, and narratives that are semantically and lexically highly similar to the originals, using only a query and examples of relevant and not relevant documents. This means that it is possible to synthesize topics that include topic titles, descriptions, and narratives that are sufficiently close to the user's intent and cognitive state in relation to their information needs (cf. the formalization in Section 3) to be useful as a basis for LLM relevance judgments.

\section{Conclusion}
In summary, we synthetically formalized information needs with LLMs and we have shown that the synthetic formalization is useful for LLM relevance assessments to make the resulting corpora more reliable. The formalization reduces the positivity bias of LLMs and improves their agreement, which, combined, yields more reliable evaluations as the subjectivity of the judgment process and the ability to evaluate new retrieval systems on a corpus is improved. When no human formalizations, such as TREC topics, are available, topics can be effectively synthesized with LLMs, even when the same LLM is also used for the relevance assessment. For this synthesis, the original query is the most important context. LLM relevance assessments that use topics synthesized from the query, relevant and non-relevant documents, align more closely with the gold-standard TREC labels. When no human topics exist in the first place, as in TREC Deep Learning, synthetically formalizing topics outperforms LLM judgments that rely solely on the original query. A leave-one-group-out evaluation showed that formalized topics improve the reliability of evaluation corpora that have LLM relevance assessments, even when only a single additional topic field, like a narrative, is used. 
While the classic title-description-narrative structure is established for human assessors, these findings suggest LLMs may also benefit from alternative topic structures that future work could explore. We observed larger variance across the different prompts used to synthetically formalize the topics compared to the variance across LLMs. This highlights the potential for future works that explore other prompt variants and context selection.

Recalling the definition introduced in Section~\ref{sec:problem}, about the relation between the information need $I$ and its formalization, the topic $T$, it is important to highlight a critical limitation: the formalization is not invertible. While the forward generative process $I \to T$ follows the natural causal direction, the inverse inference, deriving the specific information need from a topic, is mathematically indeterminate. The operational similarity of relevance judgments does not imply the similarity of the inferred information needs:
\begin{equation}
	\mathcal{R}(\hat{T}, \mathcal{D}) \approx \mathcal{R}(T, \mathcal{D}) \nRightarrow P(I \mid \hat{T}) \approx P(I \mid T)
\end{equation}

This discrepancy arises because a single information need can be expressed through vastly different topic formulations that vary not only in the set of formalized features but more importantly in ambiguity and style. For example, an oracle $O$ may produce identical judgment sets whether the topic is sampled as a sparse list of keywords or a scientifically precise statement.
While this definition is sufficient for the prevalent use case of improving automated relevance judgments, it does not allow us to conclude on the general quality of the synthetic topics or their usefulness for other applications, such as the study of individual searchers. To assess the general quality of synthetic topics, future work must develop foundational quality criteria for formalizing information needs.

In this work, we entirely relied on a topic structure established over many years of TREC experiments. However, the results showed that not all topic features were important for an improved benchmark robustness. Further, significant differences in label alignment between prompts were observed, while the similarity of the topic texts differed only marginally. This raises the question of what topic structure is best suited for LLM relevance assessors.

Automated relevance judgments reduce cost and enable scaling test collection construction. This allows for evaluating systems across other and more diverse scenarios, including specific search settings or information needs from the long tail. Furthermore, being able to accurately synthesize general-purpose topics could enable new directions for evaluations, such as evaluations that simulate information needs of specific user groups or even personalized evaluations based on user profiles as context.

\begin{acks}
	This work was supported by Deutsche Forschungsgemeinschaft
	(STELLA - 407518790 and PIXLS – 492813820), Klaus Tschira Stiftung (JoIE – 00.003.2020), and the German Federal Ministry of Education and Research~(BMBF) and Joint Science Conference~(GWK) in the PLan\_CV project (03FHP109).
	We thank Paul Thomas for providing the original GPT-4.1 judgments~\cite{thomasLargeLanguageModels2024}.

	The authors acknowledge the peoples of the Woi Wurrung and Boon Wurrung language groups of the eastern Kulin Nation on whose unceded lands ACM SIGIR 2026 was hosted. We pay our respects to their Elders past and present, and extend that respect to all Aboriginal and Torres Strait Islander peoples today and their continuing connection to land, sea, sky, and community.
\end{acks}

\bibliographystyle{splncs04}
\bibliography{conf26-gen-topic-lit}
\balance
\end{document}